\documentclass{ws-procs975x65}
\usepackage{amsmath,amsfonts,amssymb,latexsym,mathrsfs}
\usepackage{graphicx}
\usepackage{color}
\usepackage{fancybox}
\def\beq{\begin{equation}}
\def\eeq{\end{equation}}

\begin{document}

\title{The large-scale structure of the ambient boundary}
\author{Ignatios Antoniadis}

\address{LPTHE, UMR CNRS 7589, Sorbonne Universit\'es, UPMC Paris 6,\\
75005 Paris, France\\
E-mail: antoniad@lpthe.jussieu.fr\\
and\\
Albert Einstein Center for Fundamental Physics, ITP,
Bern University,\\ Sidlerstrasse 5 CH-3012 Bern, Switzerland}

\author{Spiros Cotsakis$^*$}

\address{GEODYSYC, University of the Aegean\\
Karlovassi, 83200, Samos, Greece\\
$^*$E-mail: skot@aegean.gr\\
and\\
Department of Mathematics,
American University of the Middle East\\P.O.Box 220 Dasman, 15453, Kuwait}

\begin{abstract}
This article gives a review of a recent construction, the ambient cosmological metric, and its implications for the global geometry of the universe. According to this proposal, the universe is a bounding hypersurface carrying a conformal structure and lying at the (conformal) infinity of a new, 5-geometry which satisfies the Einstein equations with fluid sources. We discuss our main results about the implied topological nature of the conformal infinity of the ambient metric, the non-existence of spacetime singularities on the boundary, as well as the validity of cosmic censorship as a direct consequence of a new principle on the boundary. 
\end{abstract}


\bodymatter


\section{Introduction}
In this paper, we review various aspects and implications of the following result developed in  Refs.~\refcite{ac1,ac2}:
\begin{theorem}
Let $(M,[g_4])$ be  a 4-dimensional spacetime with a conformal structure. Any  4-metric $g\in [g_4]$  has an ambient 5-metric $g_+$ on spacetime $V=M\times\mathbb{R}$ such that:
 \begin{itemize}
\item It satisfies the 5-dimensional Einstein equations with a fluid source on $V$
\item  $V$ has $M$ as its conformal infinity, $(\mathscr{I}_{g_{+}},\mathring{g}|_M)$
\item Any two conformally related 4-metrics on $M$, $g_1=\Omega^2g_2$,   have ambient metrics  differing  by $\mathring{g}_1|_M(0)-\mathring{g}_2|_M(0)=g_1$. Hence, $\mathscr{I}_{g_{+}}$ has a homothetic symmetry, $\mathring{g}|_M=cg_4$
 \end{itemize}
\end{theorem}
To appreciate this result, we need several distinct pieces of background, namely, various results from $AdS_5/CFT_4$ geometry, and more generally the ambient construction in conformal geometry, as well as various asymptotic limits of braneworlds. 

Witten in his work [\refcite{w98}] gives various proofs and implications of the following basic result: The $4-$dimensional Minkowski spacetime $\mathcal{M}_4$ is the boundary of $AdS_5$. Basically  one may proceed from the Poincar\'e metric $g_{AdS_5}=4(1-|y|^2)^{-2} g_E$,  to a new metric $\mathring{g}=\Omega^2 g_{AdS_5}$, and then restrict $\mathring{g}|_{S^4}$ to finally get a conformal structure on the boundary spacetime.

We now consider the \emph{inverse problem}: Starting from a conformal space-time manifold $(M,[g_4])$, is there a  metric on $V$ such that when we perform the construction we get the given conformal structure that we started with? This is the problem that occupied the fundamental work Ref.~\refcite{fg}, concerned with the construction of conformal invariants. It was shown in  Ref.~\refcite{fg} that there exists a well-defined \emph{ambient metric} (this is the Fefferman-Graham metric) $g_+$ on $M\times\mathbb{R}$ (points $(x^\mu,y)$) with the following properties (cf. Ref.~\refcite{fg}):
\begin{itemize}
  \item Locally around $M\times \{0\}$ in $M\times\mathbb{R}$, there is a smooth (non-unique) function $\Omega$ with $\Omega>0$ on $V$, $\Omega=0$ on $M$, and such that $\Omega^2g_+$ extends smoothly on $V$.
  \item $(\Omega^2g_{+})|_{TM}$ is non-degenerate on $M$ (that is its signature remains $(-++++)$ on $M$).
  \item $(\Omega^2g_{+})|_{TM}\in [g_4]$. ($M$ is the conformal infinity of $V$.)
  \item $g_+$ satisfies the Einstein equations with a cosmological constant $\Lambda$ to infinite order on $M$.
  \item $g_+$ is in normal form with respect to $g_4$:\[ g_+=y^{-2}(g_y+dy^2).\] Here, $g_y$ stands for a suitable formal power series with $g_0=g_4$. (We may also use $y$ as $\Omega$.)
  \item $g_+$ is unique: Given any two ambient metrics $g_+^1,g_+^2$ for  $(M,[g_4])$, their difference $g_+^1-g_+^2$ vanishes to infinite order everywhere along $M\times\{0\}$.
\end{itemize}

It is possible use these results in order to reveal certain drawbacks of braneworld geometry, especially important in questions of an asymptotic nature.  We know how to obtain a complete profile of all asymptotic situations that emerge when we have a bulk 5-geometry $(V,g_5)$ containing an embedded 4-dimensional braneworld $(M,g_4)$ that is either a 4-dimensional  Minkowski, or de Sitter, or Anti-de Sitter spacetime, cf. [\refcite{ack}]. In general, all asymptotic solutions have a form dictated by the method of asymptotic splittings [\refcite{skot}]. 
From these results, it is not difficult to conclude that the following properties   apply in fact to a great variety of different models of braneworlds (cf. e.g.,  Refs.~[\refcite{rs1}]-[\refcite{Forste2}] and refs. therein):
   \begin{itemize}
     \item The properties of the metric $g_4$ do not follow from those of the bulk metric $g_5$ but are dictated by field equations valid on the 4-`brane' itself.
  \item There is no conformal infinity for the 5-dimensional geometry (the brane is certainly  a kind of boundary to the bulk, but it can never be a \emph{conformal} boundary).
  \item No holographic interpretation is possible and there is no way to realize a boundary \textsc{CFT}.
\end{itemize}
In what follows, we present a novel approach in which all of the above difficulties are absent. For more details and developments, the reader is advised to look at Refs. [\refcite{ac1,ac2}].

\section{Ambient cosmology}
In this Section, we present a brief summary of the main points of our construction to produce a situation where  generic spacetimes will end up having improved properties over those we may meet in the theory of hypersurfaces in general relativity (or in its higher-dimensional extensions as above). In the proposal below, a new bounding hypersurface, the conformal infinity  of a new cosmological metric in 5-dimensional `ambient' space will be the result.

Our construction is generally one belonging to conformal geometry (cf. Ref. [\refcite{pen86}]), and may be summarized as follows (cf. [\refcite{ac1}] for more discussion and complete proofs).
\begin{enumerate}
\item Take a 4-dimensional, non-degenerate `initial'  metric $g_{\textrm{\textsc{in}}}(x^\mu)$ on spacetime $M$. This step essentially involves the Penrose conformal method.
\item Conformally deform $g_{\textrm{\textsc{in}}}$ to a new metric $g_4=\Omega^2g_{\textrm{\textsc{in}}}$  by choosing a suitable conformal factor $\Omega$. This step connects the `bad' metric $g_{\textrm{\textsc{in}}}$ with the `nice', non-degenerate, and non-singular  metric $g_4(x^\mu)$.
\item Using the method of asymptotic splittings for the 5-dimensional Einstein equations with an arbitrary (with respect to the fluid parameter $\gamma$) fluid,  solve for the 5-dimensional metric $g_5=a^2(y)g_4+dy^2$ and the matter density  $\rho_5$.
\item Transform the solutions of step 3 to suitable factored forms of the general type, (divegent part) $\times$ (smooth part).
\item  Construct the `ambient' metric in normal form, $g_+$, for the 5-dimensional Einstein equations with a fluid. This is given by the following form,
\[g_+=w^{-n}\left(\sigma^2(w)g_4(x^\mu)+dw^2\right),\] $n\in\mathbb{Q}^+$, as $w\rightarrow 0$, with $\sigma(w)$ a smooth (infinitely differentiable) function such that $\sigma(0)$ is a nonzero constant.
\item $(M,[g_4])$ is the conformal infinity of $(V,g_+)$, that is   $\mathscr{I}=\partial V=M$.
\item The metric $g_+$ is conformally compact. This means that a suitable metric $\mathring{g}$ constructed from $g_+$ extends smoothly to $V$, and its restriction to $M$, $\mathring{g}|_M$, is non-degenerate (i.e., maintains the same signature also on $M$).
\item The conformal infinity $M$ of the ambient metric $g_+$ of any metric in the conformal class $[g_4]$ is controlled by the behaviour of a constant rescaling of the `nice' metric $g_4$.
\end{enumerate}
However, uniqueness of the ambient cosmological metric is not achieved like in the Fefferman-Graham construction [\refcite{fg}]. Instead,  we find [\refcite{ac1}] an \emph{asymptotic condition} valid on the conformal infinity of the ambient space after taking suitable limits of the various possible geometric asymptotics of the problem. This method lies in the heart of the whole construction, and  is treated briefly below.

For any two conformally related 4-metrics $g_1,g_2$ in the conformal geometry of $M$, $g_1$ being the `good' (roughly meaning `regular') and $g_2$ the `bad' metric on the boundary, their ambient metrics $\mathring{g}_1|_M,\mathring{g}_2|_M$ differ by a homothetic transformation,
\beq\label{asymptotic}
\mathring{g}_2|_M(0)=c\,\mathring{g}_1|_M(0),\quad c:\;\textrm{const.}
\eeq

\textsc{Conclusion:} Starting from a conformal geometry on the spacetime $M$, the ambient cosmological metric returns a 4-geometry on  $M$ (its conformal infinity metric $\mathring{g}|_M$) that has a homothetic symmetry.
\section{Implications}
According to our proposal, our 4-dimensional world is the conformal infinity of the ambient 5-space discussed above. What are the basic implications of this proposal? We may summarize two of them as follows.
\begin{enumerate}
\item\label{1} As a conformal manifold, $(M,[g_4])$ can have no singularities.
\item Cosmic censorship on $(M,[g_4])$ is equivalent to the validity of ambient 5-metric construction,  the asymptotic condition satisfied by  the ambient metric $\mathring{g}|_M$.
    \end{enumerate}
Below, we treat (\ref{1}) in some detail, and refer to the relevant literature about the rest.

One notion that plays a key role in many theorems in global causal structure and the singularity thorems in general relativity is the convergence of a sequence of causal curves. Looking carefully at the proofs of various such results, we note the following (cf. [\refcite{ac2}] for a more complete discussion of this).
\begin{itemize}
  \item For the convergence of a sequence of causal curves to a limit curve, one uses in an essential way the Euclidean balls with their Euclidean metric and their compactness in order to extract the necessary limits.
  \item Since the Zeeman topology, unique for the conformal boundary, is strictly finer than the Euclidean metric topology, such sequences will be Zeno sequences and their convergence in the Euclidean topology will not guarantee the existence of a limit curve in the Zeeman topology.

\end{itemize}
The non-convergence of sequences of causal curves has the following implication, cf. [\refcite{ac2}].
\begin{itemize}
  \item In the proofs of the singularity theorems,  a contradiction appears when assuming the existence of a curve of length greater than some maximum starting from a spacelike Cauchy surface  $\Sigma$ (on which the mean curvature is negative) downwards to the past.
      \item One extracts a limit curve $\gamma$ (which locally maximizes the length between $\Sigma$ and an event $p$), and no curve can have length greater than that of $\gamma$.
      \item Here we cannot extract such a limit.
\end{itemize}
This result and more elaborate work  [\refcite{acp}] along these lines, opens the way for  the construction of complete spacetimes as the conformal infinities of physical theories in higher dimensional ambient space.


\begin{thebibliography}{0}

\bibitem{ac1} I. Antoniadis, S. Cotsakis, \emph{Eur. Phys. J. C. } \textbf{75:35} (2015) 1-12,  \textsc{CERN-PH-TH}/2014-176, \texttt{arXiv:1409.2220}.
\bibitem{ac2} I. Antoniadis, S. Cotsakis, \emph{Mod. Phys. Lett. A,} \textbf{30} (2015) 1550161, \texttt{arXiv:1505.04737}.
\bibitem{w98} E. Witten, \emph{Adv. Theor. Math. Phys.} \textbf{2} (1998) 253-291.
\bibitem{fg} C. Fefferman and C. Robin Graham, \emph{The Ambient Metric} (Princeton University Press, 2012), arXiv:0710.0919.
\bibitem{ack} I. Antoniadis, S. Cotsakis and I. Klaoudatou, In Proceedings of the
MG11 Meeting, H. Kleinert, R. J. Jantzen and R. Ruffini, (World
Scientific, 2008), pp. 2054-6, \texttt{arXiv:gr-qc/0701033}; \emph{Class. Quant. Grav.}\textbf{ 27} (2010) 235018, \texttt{arXiv:1010.6175};
\emph{ Fortschr. Phys.} \textbf{61} (2013) 20-49, \texttt{arXiv:1206.0090};
\emph{ Eur. Phys. J. C. } \textbf{74} (2014) 3192,  \texttt{arXiv:1406.0611}.
\bibitem{skot}
S. Cotsakis, J. D. Barrow, \emph{J. Phys. Conf. Series} \textbf{68} (2007) 012004, \texttt{arXiv:gr-qc/0608137}.
\bibitem{rs1}L. Randall and R. Sundrum, \emph{Phys. Rev. Lett.} \textbf{83}, 3370 (1999).
\bibitem{rs2}L. Randall and R. Sundrum, \emph{Phys. Rev. Lett.} \textbf{83}, 4690 (1999).

\bibitem{nima}
N. Arkani-Hamed, S. Dimopoulos, N. Kaloper, R. Sundrum,  \emph{Phys. Lett. B} \textbf{480}
(2000) 193-199;\texttt{ arXiv:hep-th/0001197}.

\bibitem{silver} S. Kachru, M. Schulz, E. Silverstein,
\emph{Phys. Rev. D} \textbf{62} (2000) 085003; \texttt{arXiv:hep-th/0002121}.

\bibitem{Gubser}  S.~S.~Gubser,   \emph{Adv. Theor. Math. Phys.} \textbf{4} (2000) 679-745; arXiv:hep-th/0002160.

\bibitem{Forste}
S.~Forste, Z.~Lalak, S.~Lavignac and H.~P.~Nilles,   \emph{Phys. Lett. B} \textbf{481} (2000) 360, \texttt{arXiv:hep-th/0002164};
  \emph{JHEP}\textbf{ 0009} (2000) 034,   \texttt{arXiv:hep-th/0006139.}

\bibitem{Forste2}
  S.~Forste, H.~P.~Nilles and I.~Zavala,
 \emph{ JCAP }{\bf 1107} (2011) 007; \texttt{arXiv:hep-th/1104.2570}.

\bibitem{pen86}R. Penrose and W. Rindler, \emph{Spinors and space-time,} Vol. 2 (Cambridge University Press, 1986).


\bibitem{acp}I. Antoniadis, S. Cotsakis and K. Papadopoulos, Mod. Phys. Lett. A31 (2016) 1650122, arXiv:1603.08117..
\end{thebibliography}
\end{document}